# Molecule-based coherent light-spin interfaces for quantum information processing – optical spin state polarization in a binuclear Europium complex

Kuppusamy Senthil Kumar,[1,2]*† Diana Serrano,[3]*† Aline M. Nonat,[4] Benoît Heinrich,[1] Lydia Karmazin,[5] Loïc J. Charbonnière,[4] Philippe Goldner,[3]* and Mario Ruben[1,2,6]*

[1] Institut de Physique et Chimie des Matériaux de Strasbourg (IPCMS), CNRS-Université de Strasbourg, 23, rue du Loess, BP 43, 67034 Strasbourg cedex 2, France.

[2] Institute of Nanotechnology, Karlsruhe Institute of Technology (KIT), Hermann-von-Helmholtz-Platz 1, 76344, Eggenstein-Leopoldshafen, Germany.

[3] Institut de Recherche de Chimie Paris (IRCP), Université PSL, Chimie ParisTech, CNRS, 75005 Paris, France.

[4] Equipe de Synthèse pour l'Analyse, IPHC, UMR 7178, CNRS-Université de Strasbourg, ECPM, 25 rue Becquerel, 67087 Strasbourg Cedex, France.

[5] Service de Radiocristallographie, Fédération de Chimie Le Bel FR2010 CNRS-Université de Strasbourg, 1 rue Blaise Pascal, BP 296/R8, 67008 Strasbourg cedex, France.

[6] Institute for Quantum Materials and Technologies (IQMT), Karlsruhe Institute of Technology (KIT), Hermann-von-Helmholtz-Platz 1, 76344, Eggenstein-Leopoldshafen, Germany.

†These two authors have equally contributed.

*e-mail: senthil.kuppusamy2@kit.edu; diana.serrano@chimieparistech.psl.eu; philippe.goldner@chimieparistech.psl.eu; mario.ruben@kit.edu

**Abstract**

The success of the emerging field of solid-state optical quantum information processing (QIP) critically depends on the access to resonant optical materials. Rare-earth ions (REIs) are suitable candidates for QIP protocols due to their extraordinary photo-physical and magnetic quantum properties such as long optical and spin coherence lifetimes ($T_2$). However, molecules incorporating REIs, despite having advantageous properties such as atomically exact quantum tunability, inherent scalability, and large portability, have not yet been studied for QIP applications. As a first testimony of the usefulness of REI molecules for optical QIP applications, we demonstrate in this study that narrow spectral holes can be burned in the inhomogeneously broadened $^5D_0 \rightarrow {^7F_0}$ optical transition of a binuclear Eu(III) complex, rendering a homogeneous linewidth ($\Gamma_h$) = 22 ± 1 MHz, which translates as $T_2$ = 14.5 ± 0.7 ns at 1.4 K. Moreover, long-lived spectral holes are observed, demonstrating efficient polarization of Eu(III) ground state nuclear spins, a fundamental requirement for all-optical spin initialization and addressing. These results elucidate the usefulness of REI-based molecular complexes as versatile coherent light-spin interfaces for applications in quantum communications and processing.

**Main**

Quantum information processing (QIP) schemes such as quantum-computing, -storage, and –communication use the quantum nature of materials to process and manipulate information[1–9]. In QIP, a dramatic improvement in computation time and secure data transmission can be achieved by creating superposition states with long coherence lifetimes ($T_2$)[10,11]. Environmental fluctuations such as lattice phonons, molecular vibrations, and magnetic moments of surrounding ions reduce the coherence lifetime of a superposition state[12,13]. To achieve superposition states with coherence lifetimes suitable for realistic applications, the coherent states must be isolated from the environmental fluctuations. Optically active impurity systems, for example, colour centers in diamond[14] and rare-earth ions (REI) embedded in a suitable matrix[3,8,15–24], have been reported to act as ideal materials to achieve superposition states with long coherence lifetimes. Among the impurity systems studied, REIs are well-suited for QIP applications due to the following intrinsic properties, (i) long optical coherence lifetimes due to the well-shielded nature of the 4f-4f optical transitions from the surrounding environment by the outer 5s and 6p closed orbitals, and (ii) the presence of nuclear spins (*I*), which enables the creation of nuclear spin superposition states with long coherence lifetimes, useful for storing quantum states[25–27]. Importantly, the exceptionally good optical coherence lifetimes associated with the 4f-4f optical transitions, covering the whole visible and IR spectral range, allow for coherent optical addressing and manipulation of nuclear spin states, rendering REIs useful for QIP applications[28–33].

Non-Kramers rare-earth ions with even number of f-electrons, for example, Eu(III), Pr(III), or Tm(III), embedded in a matrix with low average magnetic moments have been extensively studied for the implementation of QIP schemes[34]. The $^5D_0 \rightarrow {^7F_0}$ transition of Eu(III) is of particular interest, because the $^5D_0 \rightarrow {^7F_0}$ transition is an induced electric dipole transition[35] and is largely unaffected by the magnetic field fluctuations arising from the surrounding environment, thereby long optical coherence lifetimes are associated with the transition. The QIP utility of Eu(III)-doped ceramics, powders, crystals, and nanoparticles—featuring $^5D_0 \rightarrow {^7F_0}$ transition—has been elucidated in the literature[3,15,18,19,36,37]. However, it is difficult to tailor-make such systems with desirable optical properties by means of chemical and physical manipulations, limiting the utility of such materials for QIP applications. On the other hand, optical properties of molecular Eu(III) complexes can be easily tuned by ligand field- and molecular energy level engineering approaches. Further, by synthesizing isotopically enriched nuclear spin-free ligands, minimization of multiphonon relaxation pathways could be realized, thereby high quantum efficiencies and narrow inhomogeneous broadening of $^5D_0 \rightarrow {^7F_0}$ transition could be achieved. Importantly, Eu(III) complexes with isotopically pure emitting centers, for example, $^{151}$Eu or $^{153}$Eu, both with $I$ = 5/2, can be synthesized via isotopologues coordination chemistry[38,39], enabling the use of nuclear spin states as storage states for QIP applications.

Frequency domain techniques such as transient spectral hole burning (SHB) have been used to optically probe superposition state lifetimes in REI systems, for example, Eu(III):$Y_2O_3$[36]. Transient spectral holes appear as dips in absorption or excitation spectra when a portion of active ions do not show ground state absorption anymore after optical pumping and for a certain time duration. The utility of transient SHB for measuring coherence lifetimes relies on the following factors, (i) existence of inhomogeneously broadened transitions with $\Gamma_{inh}$ >> $\Gamma_h$, where $\Gamma_{inh}$ and $\Gamma_h$ are the optical inhomogeneous and homogeneous linewidths, respectively; (ii) $\Gamma_h$ >> $\Gamma_{laser}$, where $\Gamma_{laser}$ is the excitation laser linewidth; and (iii) $\Gamma_h < \tau^{-1}$ and/or $T_1^{-1}$, for SHB in the excited-state and/or in the ground-state spin levels, where $\tau$ and $T_1$ are the excited-state and spin lifetimes, respectively. In these conditions, the hole width ($\Gamma_{hole}$) is equal to $2\Gamma_h$, with $\Gamma_h$ inversely proportional to the optical coherence lifetime ($\Gamma_h$ = $1/\pi T_2$). Reports on SHB in REIs are rather limited to REIs dispersed in matrices such as $Er^{3+}$:$^7LiYF_4$, $Nd^{3+}$:$YVO_4$, $Eu^{3+}$: $Y_2SiO_5$ and $Eu^{3+}$:$Y_2O_3$[34].

To the best of our knowledge, transient SHB in a molecular REI system—especially in an Eu(III) complex—is yet to be reported. In this study, we demonstrate—for the first time—transient SHB (see methods, section S1, and Fig. S1-S3 for experimental details) in the inhomogeneously broadened $^5D_0 \rightarrow {^7F_0}$ optical transition associated with a binuclear molecular Eu(III) complex—**[Eu$_2$Cl$_6$(4-picNO)$_4$(μ$_2$-4-picNO)$_2$]·2H$_2$O**—hereafter **[Eu$_2$]**. The measured hole width yields a homogeneous linewidth of 22 ± 1 MHz, which corresponds to an optical T$_2$ of 14.5 ± 0.7 ns. A hole decay time of 2.1 s is observed, consistent with nuclear spin relaxation. This confirms the ability of the SHB method used here to optically polarize the nuclear spin states in the **[Eu$_2$]** complex. The results presented in this study might open a new avenue in quantum materials research direction by placing the focus on REI-based molecular materials.

**Results and discussion**

**Preparation and X-ray structure analysis of [Eu$_2$]**

The preparation of **[Eu$_2$]** was performed by treating the commercially available EuCl$_3$·6H$_2$O and 4-picoline-N-Oxide (4-picNO) in 1:3 ratio in water followed by recrystallization of the crude reaction mixture from ethanol-ethyl acetate solvent mixture (Scheme S1). X-ray crystallographic analysis of the complex crystals revealed binuclear structure, as shown in Fig. 1a and b; see Fig. S4 for packing in the crystal lattice. The complex crystallized in the centrosymmetric $P\bar{1}$ space group, belonging to the triclinic crystal system.

The neutral Eu(III)-dimer is composed of six 4-picNO ligands and six chloride ligands. The O$_4$Cl$_3$ coordination environment around each Eu(III) ion is best described as pentagonal bipyramidal with a continuous shape measure (CShM)[40] of 1.513, as shown in Fig. 1b. In **[Eu$_2$]**, the equatorial positions (edges) of each pentagonal bipyramid are occupied by two monodentate 4-picNO ligands in trans fashion, one chloride ligand, and two μ$_2$-4-picNO ligands. The axial positions of each pentagonal bipyramid are occupied by the remaining chloride ligands. The intramolecular Eu···Eu distance is 4.273(4) Å. A good match between the powder (PXRD) and single crystal (SCXRD) patterns (Fig. S5) and the similar unit cell parameters (Table S1) obtained from the indexing of the PXRD and SCXRD data unambiguously prove the phase purity of the crystalline material utilized for photophysical and SHB studies.

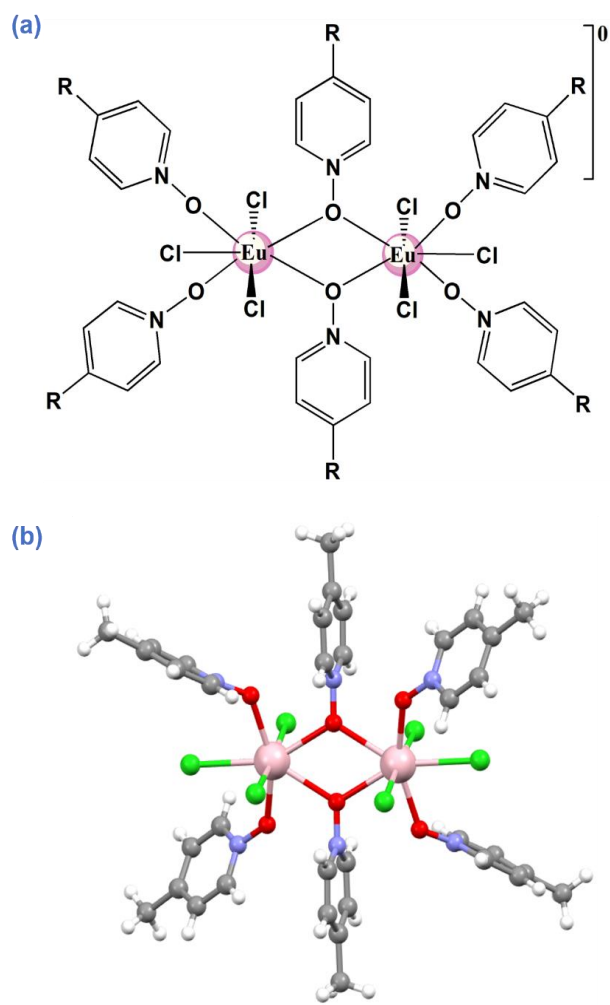

**Fig. 1 | Structure of [Eu$_2$Cl$_6$(4-picNO)$_4$(μ$_2$-4-picNO)$_2$]·2H$_2$O. (a)** Molecular structure of the complex showing the ligands (R = CH$_3$) coordinating with the Eu(III) centers. The complex was prepared by treating EuCl$_3$·6H$_2$O with **4-picNO** ligand dissolved in water followed by a recrystallization step from ethanol (EtOH)/ethyl acetate (EtOAc) solvent mixture. **(b)** X-ray crystal structure of the complex. Coordination geometry around each Eu(III) centre of the complex is best described as pentagonal bipyramidal. The co-crystallized water molecules are omitted for clarity. Colour code: H, white; C, grey; N, blue; Cl, green; Eu, pink. Crystal Data: C$_{36}$H$_{46}$N$_6$O$_8$Cl$_6$Eu$_2$, $M_r$ = 1207.41, triclinic, $P\bar{1}$ (No. 2), a = 9.7938(11) Å, b = 10.5255(12) Å, c = 12.9179(15) Å, α = 66.258(3)°, β = 75.915(4)°, γ = 82.000(4)°, V =

1181.0(2) Å$^3$, $T$ = 173(2) K, $Z$ = 1, $Z'$ = 0.5, $\mu$(MoK$_\alpha$) = 3.022 mm$^{-1}$, 54086 reflections measured, 6981 unique ($R_{int}$ = 0.0646), which were used in all calculations. The final $wR_2$ was 0.1383 (all data) and $R_1$ was 0.0447 (I > 2σ(I)). The crystal data were generated using Olex 2[41].

**Photophysical studies**

The photoluminescence characteristics of the complex were studied in the solid-state by exciting the 4-picNO-based transition centerd at ~330 nm. Intense Eu(III)-based line-like emission bands corresponding to the $^5D_0\rightarrow{}^7F_J$ ($J$ = 0-6) transitions were observed, as depicted in Fig. 2b and Fig. S6a. Remarkably, the $^5D_0\rightarrow{}^7F_0$ transition is sharp and non-degenerate (*vide infra*), thus indicating the presence of a single Eu(III) site, that is, the presence of the same spectroscopic site symmetry around the two Eu centers in the complex. Moreover, the occurrence of the $^5D_0\rightarrow{}^7F_0$ transition is a testimony of the low-symmetric coordination environment around the Eu(III) ions in the complex (Fig. 2b). The near infra-red (NIR) $^5D_0\rightarrow{}^7F_{5,6}$ transitions were seen around 745 nm ($^5D_0\rightarrow{}^7F_5$) and in the 801 nm to 836 nm ($^5D_0\rightarrow{}^7F_6$) region (Fig. S6a). The $^5D_0\rightarrow{}^7F_5$ and $^5D_0\rightarrow{}^7F_6$ transitions amount to 3.3% and 7.5%, respectively, of the total emission intensity. Remarkably, the $^5D_0\rightarrow{}^7F_4$ transition centered around 700 nm accounts for 26% of the total emission. Overall, a deep red emission with chromaticity coordinates (x = 0.6615, y = 0.3382) on the commission internationale de l'éclairage (CIE) color space (Fig. S6b) was observed for the complex.

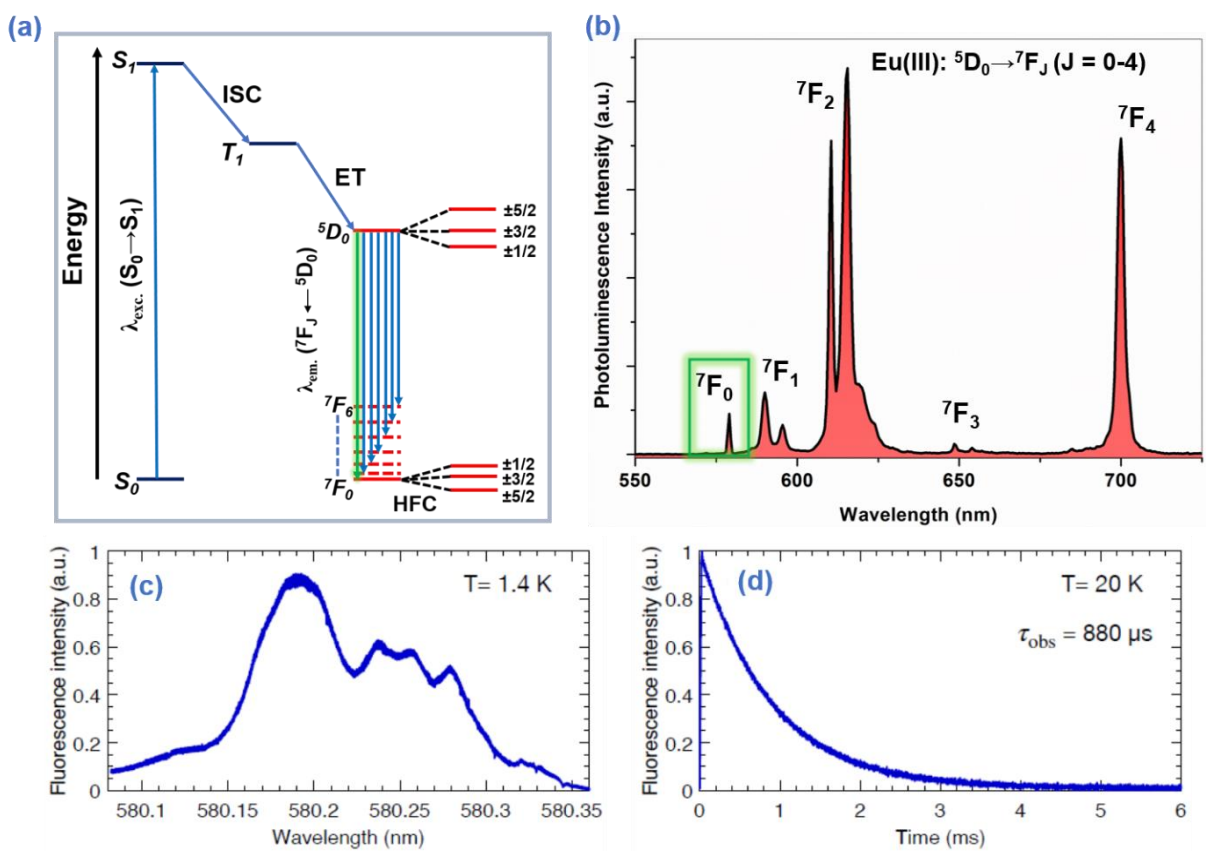

**Fig. 2 | Photophysical properties of [Eu$_2$Cl$_6$(4-picNO)$_4$($\mu_2$-4-picNO)$_2$]·2H$_2$O in the solid-state.**
**(a)** General mechanism of PL sensitization in Eu(III) complexes: The Eu(III)-based $^5D_0$ receiving level is populated after a series of excitation, intersystem crossing (ISC), and $T_1 \rightarrow {}^5D_0$ energy transfer (ET) processes. The radiative relaxation of $^5D_0$ level to ground $^7F_J$ ($J$ = 0-6) crystal field levels manifests as line-like luminescence. Among the $^5D_0 \rightarrow {}^7F_J$ transitions, the $^5D_0 \rightarrow {}^7F_0$ transition is suited for QIP applications due to its narrow linewidth and long coherence lifetimes of the hyperfine states—±5/2, ±3/2, ±1/2 (I = 5/2 for $^{151}$Eu/$^{153}$Eu) associated with the $^7F_0$ ground state level. **(b)** Photoluminescence spectrum showing the $^5D_0 \rightarrow {}^7F_J$ ($J$ = 0-4) transitions in the visible and near-IR range. **(c)** Photoluminescence excitation (PLE) spectrum of the $^5D_0 \rightarrow {}^7F_0$ transition of the complex measured at 1.4 K. An inhomogeneous linewidth ($\Gamma_{inh}$) of 50 MHz is calculated for the main peak centred at 580.185 nm (vac.) **(d)** Luminescence decay of the $^5D_0$ excited state of the Eu(III) complex measured at 20 K under resonant excitation at 580.185 nm. The fluorescence decay is well fitted by a single exponential function, suggesting a single emitting Eu(III) centre. The observed excited state lifetime ($\tau_{obs}$) of 880 μs indicates the absence of nonradiative relaxation channels, especially the ones that could originate from the lattice water molecules.

A strictly mono-exponential decay with luminescence lifetime ($\tau_{obs}$) of 822 μs, upon monitoring the luminescence decay profiles at 616 nm, was observed at RT (Fig. 6c), confirming the presence of a single emitting Eu(III) species. A total emission quantum yield ($Q_{tot}$) of 38 ± 6% ($\lambda_{exc}$ = 330 nm) was determined in the solid-state using the absolute method. The total PL quantum yield, $Q_{tot}$, can be represented as the product of the sensitization efficiency ($\eta_{sens}$) and the intrinsic quantum yield ($Q_{Eu}$) of the europium emission from the $^5D_0$ level:

$$Q_{tot} = \eta_{sens} \times Q_{Eu} = \eta_{sens} \times (\tau_{obs}/\tau_{rad}) \qquad (1)$$

Where $\tau_{rad}$ is the radiative lifetime of the $^5D_0$ level. The radiative lifetime is expressed as[42]

$$\tau_{rad} = [A_{MD} \times (n)^3 \times I_{tot}/I_{MD}]^{-1} \qquad (2)$$

Where $A_{MD}$ is the spontaneous emission probability for the $^5D_0 \rightarrow {}^7F_1$ transition in vacuum (taken here as 14.65 s$^{-1}$), n is the refractive index (taken here as 1.5) of the solid complex, and $I_{tot}/I_{MD}$ is the ratio of the total integrated emission intensity (for all transitions from the $^5D_0$ state to the $^7F_J$ manifold ($I_{tot}$)) to that of the magnetic dipole $^5D_0 \rightarrow {}^7F_1$ transition ($I_{MD}$). The values of $\tau_{rad}$ = 893 μs and $\eta_{sens}$ = 41% were calculated from equations 1 and 2. A near quantitative intrinsic Eu quantum yield ($Q_{Eu}$) = ~92% was obtained.

**Low-temperature high-resolution spectroscopy and spectral hole burning**

As mentioned in the introduction, the $^5D_0 \rightarrow ^7F_0$ transition of Eu(III) is of particular interest for QIP and optical quantum technologies. The photoluminescence excitation (PLE) spectrum, recorded at 1.4 K, of the $^5D_0 \rightarrow ^7F_0$ optical transition in the binuclear complex (Fig. 2c) is composed of several partially resolved peaks. The most prominent appeared at 580.185 nm (vac.) with a full width at half maximum (FWHM) of 0.06 nm (50 GHz or 1.7 cm$^{-1}$). This is comparable to inhomogeneous linewidths observed in Cr(III) compounds[43] and larger than optical inhomogeneous linewidths in most inorganic crystals, in the few GHz range[44]. The complexity of this absorption spectrum for a transition taking place between two singlet electronic levels, and with a single Eu(III) site in the compound, raises the question about the origin of the different peaks. Several absorption peaks observed for the Eu(III) $^5D_0 \rightarrow ^7F_0$ transition have been attributed to a combination of hyperfine structures and isotope shifts from nuclear spin containing Cl, O, and H-based nearest neighbor ligands[45]. However, the observed energy splitting (Fig. 2c) in the order of tens of GHz, that is, several cm$^{-1}$, are too large to be attributed to any of those effects here[46]. Alternatively, the side peaks on the lower energy side of the main peak could correspond to molecular rotations, presenting energies ranging from 0.1 to tens of cm$^{-1}$ [46]. However, further investigations are needed to confirm this hypothesis. The $^5D_0$ fluorescence decay measured at 20 K under resonant excitation at 580.185 nm is shown in Fig. 2d. A single exponential fit of the decay profile yielded a lifetime ($\tau_{obs}$) = 880 µs, which is slightly longer than that found at room temperature. The experimentally observed $\tau_{obs}$ is comparable to the estimated $\tau_{rad}$ associated with the complex.

Homogeneous linewidths are of major importance to evaluate the potential of a system for optical QIP applications. In particular, narrow homogeneous linewidths are required for optical quantum storage and subsequent quantum coherence transfer to the nuclear spin states. Homogeneous linewidth values are, however, difficult to access from the inhomogeneously broadened lines of REI-based systems by classical absorption and fluorescence techniques. To circumvent this issue, we have used SHB[47] and probed the optical homogeneous linewidth of $^5D_0 \rightarrow {}^7F_0$ transition associated with **[Eu$_2$]** and demonstrated selective optical addressing of the ground-state nuclear spin levels in the complex (see methods, section S1, and Fig. S1-S3 for details). The applied optical pumping continuously transfers a resonant ensemble of Eu(III) molecules to the excited state. Once in the excited state, three decay channels are possible, because Eu(III) presents three doubly-degenerate ground-state nuclear spin levels (±5/2, ±3/2, and ±1/2; for $^{151/153}$Eu, I = 5/2) (Fig. 3b). Ions decaying back to the initial state will be immediately pumped back to the excited state. In contrast, those relaxing to a different nuclear spin level from the original one are no longer resonant with the excitation laser, therefore, population remains there until spin relaxation occurs (Fig. 3b).

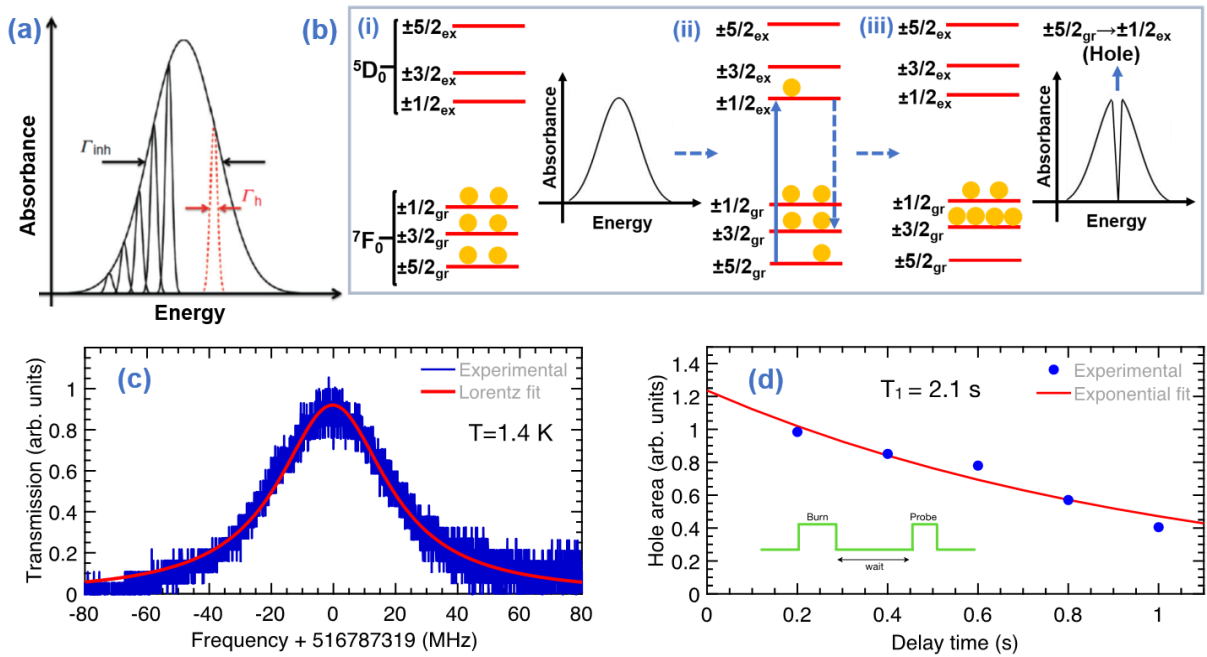

**Fig. 3 | Spectral hole burning (SHB) in the $^5D_0 \rightarrow {^7F_0}$ transition of $[Eu_2Cl_6(4\text{-picNO})_4(\mu_2\text{-4-picNO})_2] \cdot 2H_2O$. (a)** An inhomogeneously ($\Gamma_{inh}$) broadened line is composed of narrow homogeneously ($\Gamma_h$) broadened lines. Selective optical excitation of one or several homogeneously broadened lines is used to burn spectral holes with implications for QIP applications. **(b)** Simplified mechanism of SHB. In the diagram, $\pm1/2_{gr}$, $\pm3/2_{gr}$, $\pm5/2_{gr}$ and $\pm1/2_{ex}$, $\pm3/2_{ex}$, $\pm5/2_{ex}$ correspond to hyperfine levels associated with the ground $^7F_0$ and excited $^5D_0$ states, respectively. (i) A laser scan reveals an inhomogeneously broadened absorption spectrum, like the one shown in **(a)**, due to the excitation of multiple sites with infinitesimally small energy difference. (ii) Selective and continuous laser pumping transfers population from the ground $\pm5/2_{gr}$ ($^7F_0$) hyperfine level to the excited $\pm1/2_{ex}$ ($^5D_0$) hyperfine level. (iii) Population relaxation from level $\pm1/2_{ex}$ to level $\pm3/2_{gr}$, under the ambit of level $\pm3/2_{gr}$ having sufficiently longer lifetime than levels $\pm1/2_{gr}$ and $\pm5/2_{gr}$, results in a decreased population in level $\pm5/2_{gr}$, producing a hole in the inhomogeneously broadened absorption spectrum. **(c)** Spectral hole burned in the $^5D_0 \rightarrow {^7F_0}$ transition of the Eu(III) complex. A Lorentzian fit reveals an FWHM of 43 ± 2 MHz, corresponding to a homogeneous linewidth

($\Gamma_h$) of 22 ± 1 MHz. An optical coherence lifetime ($T_2$) of ~14.5 ns is calculated using the relation $\Gamma_h = (1/\pi T_2)$. **(d)** The decay of the hole area as a function of delay time before readout, showing the relaxation rate of the nuclear spin levels in the binuclear complex. Figures (a) and (b) are partially reproduced/adopted with permission from reference 34.

As the pumping proceeds, the original spin state is progressively emptied, and the population is transferred to the other spin level (Fig. 3b). Scanning the laser over the optical inhomogeneous line then reveals an increased transmission at the burning frequency, that is, a spectral hole. For an excitation laser linewidth much narrower compared to measured hole widths, the optical homogeneous linewidth of the transition can be derived from the hole width as $\Gamma_h=\Gamma_{hole}/2$. The SHB method is used for tailoring absorption profiles in several quantum storage protocols, and initialization of spin population[29,48] relies on the ground-state nuclear spin level with long population lifetimes, or at least, substantially longer than the time waited before probing the spectral hole.

A spectral hole burned in the $^5D_0\rightarrow{^7F_0}$ transition of Eu(III) is shown in Fig. 3c. The hole was fitted by a Lorentz peak function, yielding a hole width ($\Gamma_{hole}$) of 43 ± 2 MHz at FWHM, corresponding to a homogeneous linewidth of 22 ± 1 MHz; therefore, an optical coherence lifetime $T_2$ of 14.5 ± 0.7 ns is estimated. The observed $\Gamma_h$ for **[Eu₂]** is relatively large compared to Eu(III) ions dispersed in matrices[44,49]. However, the $\Gamma_h$ is about an order of magnitude lower than homogeneous linewidths observed in most diluted Cr(III) molecular systems at low temperature[43] and comparable to the ones reported for NV centers in diamond[50]. The broad $\Gamma_h$ in Cr(III) systems has been attributed to intramolecular relaxations[43]. Intramolecular relaxations related to low-energy rotations could also explain the broader homogeneous line in **[Eu₂]** with respect to inorganic crystals. Moreover, a strong sensitivity of $\Gamma_h$ to magnetic impurities surrounding the Eu(III) sites was demonstrated in Eu(III):Y$_2$O$_3$[49] and EuCl$_3$·6H$_2$O[51]. This mechanism is also to be considered here due to the direct coordination of nuclear spin containing chloride (Cl$^-$) ligands and the presence of hydrogen atoms in the **4-picNO** ligand skeleton.

To further confirm that the observed spectral holes result from population transfer to a different spin level, optical spin control manipulations were carried out. First, the determined minimum time delay (5 ms) between hole burning and hole readout pulses was much longer than excited state lifetime of 880 μs, ensuring that no population remained in the optical excited state (Fig. S2). Second, holes could be erased by scanning the laser over a frequency of 200 MHz with a high intensity. The laser pulses excite the spin population in the ground-hyperfine spilt levels, including the populations present in the storage levels, because typical Eu ground state hyperfine splittings are in the range 25-100 MHz[52]. This broad excitation leads to the reestablishment of the initial thermal equilibrium population distribution following spontaneous relaxation from the excited state. Finally, we observed that the hole could be erased and then burned again at a different frequency within the optical inhomogeneous line (Fig. S3) unambiguously elucidating selective and reversible optical addressing of a sub-ensemble of Eu(III) ions.

The decay of the hole area was observed by increasing the delay between burning and readout pulses (Fig. 3d). This measurement provides insight into the relaxation constant ($T_1$) for the nuclear spin levels of Eu(III), estimated as 2.1 s by exponential decay fit. As it is more than three orders of magnitude larger than the excited state lifetime, efficient spin population transfer can be achieved, enabling, for example, spin state initialization. We note that the spin $T_1$ does not seem limited by spin flip-flops between neighboring Eu(III) ions since the intramolecular flip-flop rate is expected between $10^{-4}$ and $10^{-5}$ s$^{-1}$ [53], which is remarkably lower than the measured relaxation rate ($T_1^{-1}$) of 0.5 s$^{-1}$.

**Conclusion**

Long-lived spectral holes were prepared in the inhomogeneously broadened $^5D_0 \rightarrow {^7F_0}$ optical transition of the binuclear molecular Eu(III) complex—**[Eu$_2$]**—at 1.4 K. The hole width corresponds to a homogeneous linewidth ($\Gamma_h$) of 22 $\pm$ 1 MHz, which is equivalent to a coherence lifetime ($T_2$) of 14.5 $\pm$ 0.7 ns. These values correspond to those typically observed for Eu(III) ions dispersed in solid matrices, and the burn mechanism is attributed to population redistribution within ground-state nuclear spin levels of Eu(III) that act as shelving levels. The observation of narrow optical inhomogeneous linewidth and nuclear spin state polarization associated with molecular **[Eu$_2$]**—reported for the first time in this study—relies on the presence of narrow $^5D_0 \rightarrow {^7F_0}$ optical transition in the complex. The usefulness of such transition for QIP applications could be enhanced adopting molecular engineering approaches. The tunable nature of molecular properties with atomic precision could provide access to quantum materials with technologically relevant spin and optical coherence lifetimes. For example, ideal Eu(III), REI-based, in general, complexes capable of showing long spin and optical coherence times could be obtained by designing ligands with deuterated backbones and nuclear spin-free donor sites and complexing such ligands with isotopically enriched REIs. Overall, the results presented in this study demonstrate the utility of molecules as coherent light-spin interfaces, useful for the realization of long-sought optical quantum technologies such as quantum memories, processors, and single photon sources.

**Methods**

Experimental descriptions on the preparation of the complex, X-ray crystallography, and powder X-ray diffraction analysis of the complex are detailed in the supporting information associated with this article.

**Steady state photoluminescence spectroscopy**

Steady state emission spectra were recorded on a FLP920 spectrometer from Edinburgh Instrument working with a continuous 450 W Xe lamp and a red sensitive Hamamatsu R928 photomultiplier in Peltier housing. The emission spectrum of the europium complex in the NIR region was also measured by using a nitrogen cooled Hamamatsu R5509-72 Vis-NIR (300 - 1700 nm) affording similar results. All spectra were corrected for the instrumental functions. When necessary, high pass filters at 330 nm, 395 nm, 455 nm, or 850 nm were used to eliminate the second order artefacts.

Phosphorescence lifetimes were measured on the same instrument working in the Multi Channels Spectroscopy (MCS) mode and using a Xenon flash lamp as the excitation source. Errors on lifetimes are ±10%. Luminescence quantum yields were measured with a G8 Integrating Sphere (GMP SA, Switzerland) according to the absolute method detailed in ref[54]. Estimated errors are ±15%.

**Low temperature high resolution spectroscopy and spectral hole burning measurements**

Photoluminescence excitation (PLE), fluorescence lifetime and spectral hole burning (SHB) measurements were carried out in a He bath cryostat (Janis SVT-200). The molecular crystals were hold, within the cryostat, by a home-built container with front and rear optical access. Temperature was regulated by acting on the liquid He volume and pressure in the sample chamber, and it was continuously monitored by a Si diode (Lakeshore DT-610). Optical excitation was carried out by a tunable CW dye laser (Sirah Matisse DS) with 250 kHz linewidth. The excitation beam was focused onto the sample and transmitted scattered light was collected by a 75-mm-diameter lens. Signals were detected with an avalanche photo diode (Thorlabs 110 A/M). The experimental setup is schematically represented in Fig. S1. PLE spectra were recorded by scanning the excitation wavelength ($\lambda_{exc}$) between 580.08 and 580.35 nm while monitoring the fluorescence intensity from the $^5D_0$ excited state at 580.185 nm (vac.). A long-pass filter (cut-off at 590 nm) was placed in front of the detector to reject the excitation wavelength. SHB spectra and fluorescence decays were recorded by modulating the CW output of the laser in frequency and amplitude with an acousto-optic modulator (AOM AA Optoelectronic MT200-B100A0, 5-VIS, 200 MHz central frequency), set in double-pass configuration, and driven by an arbitrary waveform generator with 625 MS s$^{-1}$ sampling rate (Agilent N8242A). The $^5D_0$ fluorescence decay was recorded after a single excitation pulse of 2 ms at 580.11 nm. The SHB sequence was formed of ten burning pulses of 2 ms length, with an excitation power of 30 mW, and fixed excitation wavelength of 580.19 nm. The waiting time between pulses was set to 3 ms to enable spontaneous relaxation from the excited state to the ground state nuclear spin levels before a new burn. The spectral hole was then probed by monitoring the transmission of a frequency scanning pulse of 2 ms duration, with a frequency span of 200 MHz and excitation power of 5 mW. The recorded transmission under burn conditions was corrected from transmission obtained without burning, to cancel out the frequency-dependent response of the AOM. The delay before readout was varied from a minimum of 5 ms to a maximum of 1 s. Every SHB sequence was ended by a series of high-power frequency-scanning pulses to reset the ground-state population back to equilibrium. Spectral holes were recorded after averaging 50 sequences to improve the signal to noise ratio, for a better estimation of the hole width.

**Data availability**

X-ray crystallographic data of the complex, in the form of cif file, discussed in this article can be obtained from the Cambridge Crystallographic Data Centre—CCDC 1863020.

**Acknowledgements**

M.R. thanks the grant agencies innovation FRC, for the financial support for the project Molecular *Qudits*: Isotopologoues Coordination Chemistry and the DFG priority program 1928 "COORNETS" for generous support. D. S. and P. G. have received funding from the European Union's Horizon 2020 research and innovation programme under grant agreement No 820391 (Square). A.M.N. and L.J.C. thank the French Agence National de la Recherche for financial support (Neutrino project n° ANR-16-CE09_0015-02).

**Author contributions**

M.R. and P.G. conceived and supervised the project. K.S. synthesized and characterized the complex. L.K. determined the X-ray structure of the complex. B.H. performed powder X-ray diffraction studies and indexed the patterns. A.M.N. and L.J.C. carried out the steady-state photoluminescence studies. D.S. and P.G. performed PLE and SHB measurements. K.S. D.S.

and M.R. wrote the manuscript. All the authors have read and commented on the manuscript.

**Additional information**

Supplementary information and chemical compound information are available in the online version of the paper. Reprints and permissions information is available online at www.nature.com/reprints. Correspondence and requests for materials should be addressed to M.R.

**Competing financial interests**

The authors declare no competing interests.

**Graphical abstract**

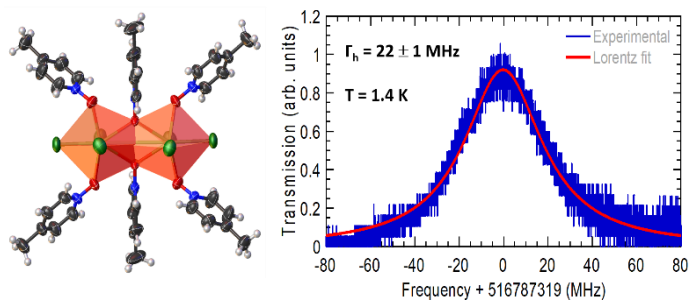

Spectral hole burning (SHB) in the inhomogeneously broadened $^5D_0 \rightarrow {}^7F_0$ transition of a binuclear rare earth ion, Eu(III), complex—**[Eu₂]**—is demonstrated for the first time, paving the way for the realization of quantum information processing (QIP) applications based on molecular materials.

# Supporting Information

# Molecule-based coherent light-spin interfaces for quantum information processing – optical spin state polarization in a binuclear Europium complex

Kuppusamy Senthil Kumar,[1,2*†] Diana Serrano,[3*†] Aline M. Nonat,[4] Benoît Heinrich,[1] Lydia Karmazin,[5] Loïc J. Charbonnière,[4] Philippe Goldner,[3*] and Mario Ruben[1,2,6*]

[1]Institut de Physique et Chimie des Matériaux de Strasbourg (IPCMS), CNRS-Université de Strasbourg, 23, rue du Loess, BP 43, 67034 Strasbourg cedex 2, France.

[2]Institute of Nanotechnology, Karlsruhe Institute of Technology (KIT), Hermann-von-Helmholtz-Platz 1, 76344, Eggenstein-Leopoldshafen, Germany.

[3]Institut de Recherche de Chimie Paris (IRCP), Université PSL, Chimie ParisTech, CNRS, 75005 Paris, France.

[4]Equipe de Synthèse pour l'Analyse, IPHC, UMR 7178, CNRS-Université de Strasbourg, ECPM, 25 rue Becquerel, 67087 Strasbourg Cedex, France.

[5]Service de Radiocristallographie, Fédération de Chimie Le Bel FR2010 CNRS-Université de Strasbourg, 1 rue Blaise Pascal, BP 296/R8, 67008 Strasbourg cedex, France.

[6]Institute for Quantum Materials and Technologies (IQMT), Karlsruhe Institute of Technology (KIT), Hermann-von-Helmholtz-Platz 1, 76344, Eggenstein-Leopoldshafen, Germany.

†These two authors have equally contributed.

*e-mail: senthil.kuppusamy2@kit.edu; diana.serrano@chimieparistech.psl.eu; philippe.goldner@chimieparistech.psl.eu; mario.ruben@kit.edu

## Contents

# S1. Experimental

## S1.1. Schematic of experimental setup used for spectral hole burning studies

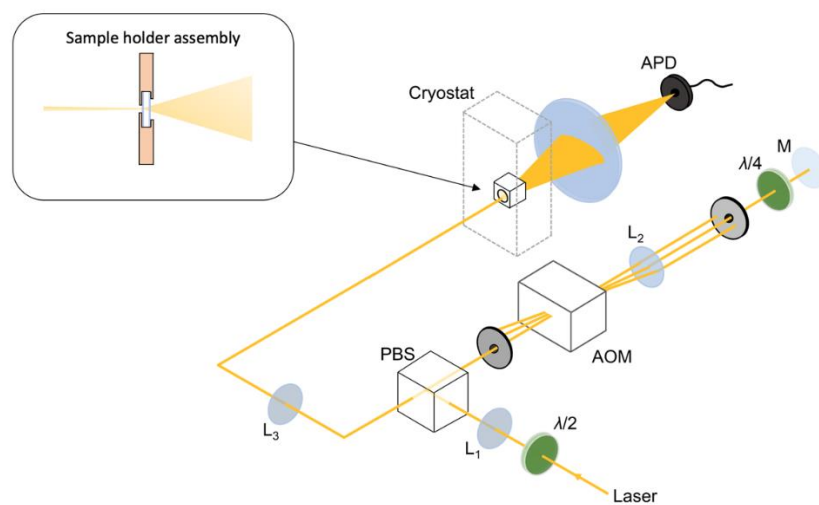

**Fig. S1 |** Experimental setup. AOM stands for acousto-optic modulator, PBS for polarizing beam splitter and APD for avalanche photo diode. $L_n$ (with n=1,2,3) are optical lenses. Unlabeled elements correspond to apertures. **Inset:** Home-built sample holder assembly. The molecular crystals are placed between two glass plates, within a brass container with front and rear apertures.

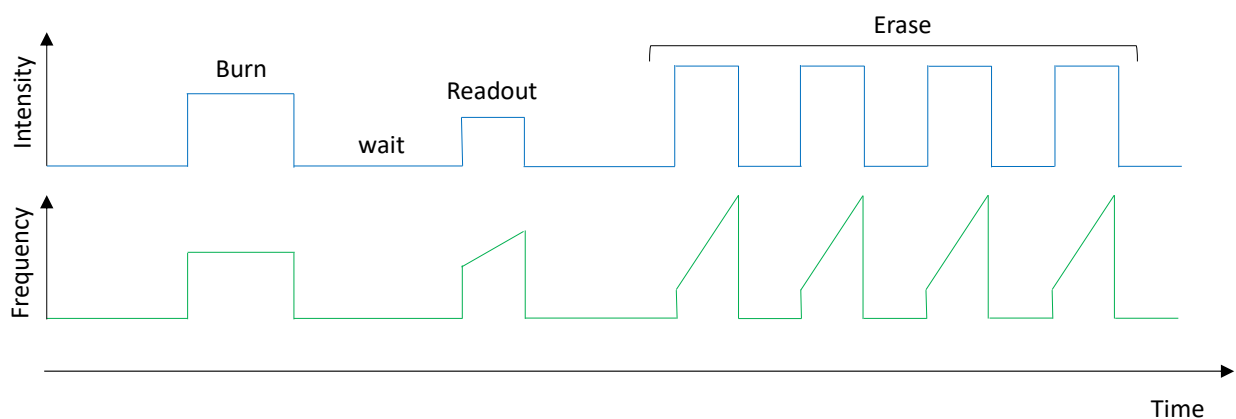

**Fig. S2 |** Spectral hole burning pulse sequence. Traces show laser intensity and frequency as a function of time. Burning was maintained for 2 ms and the waiting time before readout was varied from 5 ms to 1 s. The sequence was ended up by a series of hole erasing pulses. See experimental description for more details.

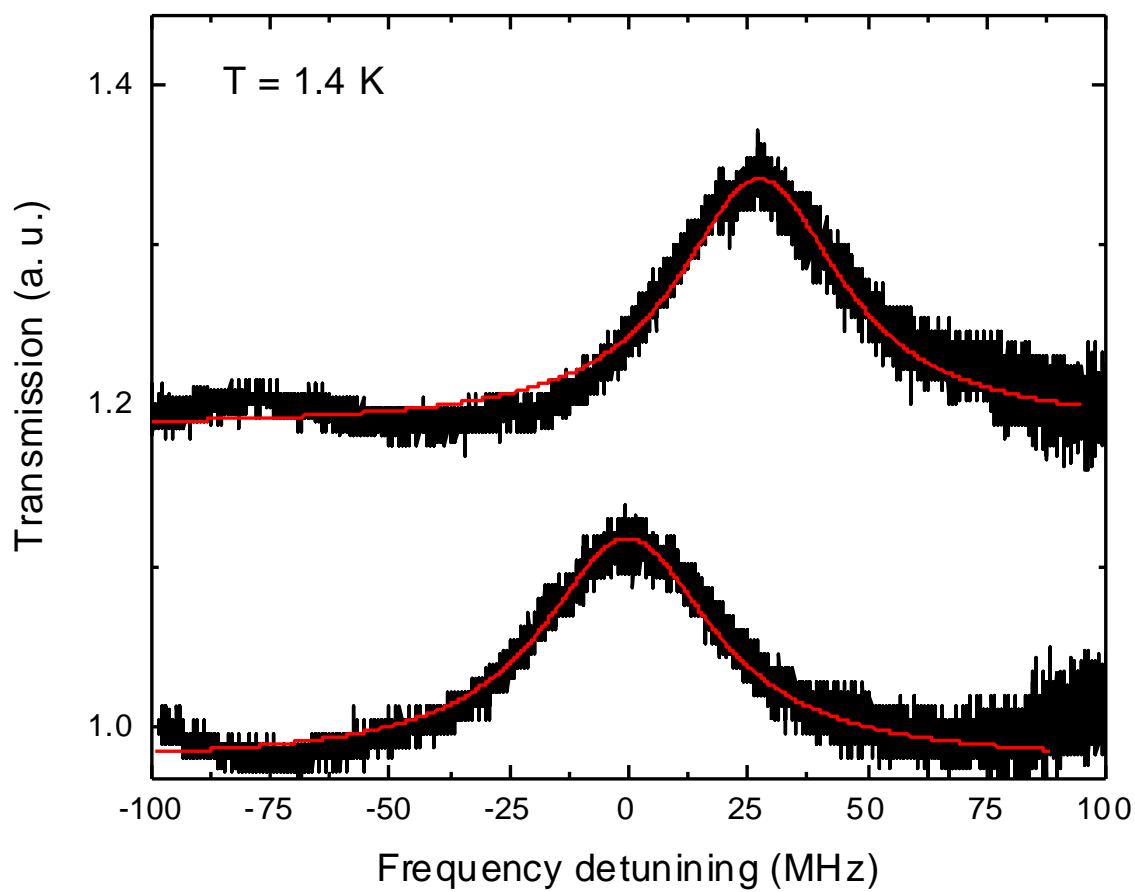

**Fig. S3 |** Spectral holes burned at two frequencies, separated by 5 MH, within the optical inhomogeneous transition. As expected, similar hole with is observed independently of the burning frequency. Data are vertically shifted for clarity.

## S1.2. Preparation of the [Eu$_2$Cl$_6$(PicNO)$_4$(µ$_2$-PicNO)$_2$]·2H$_2$O (**[Eu$_2$]**)

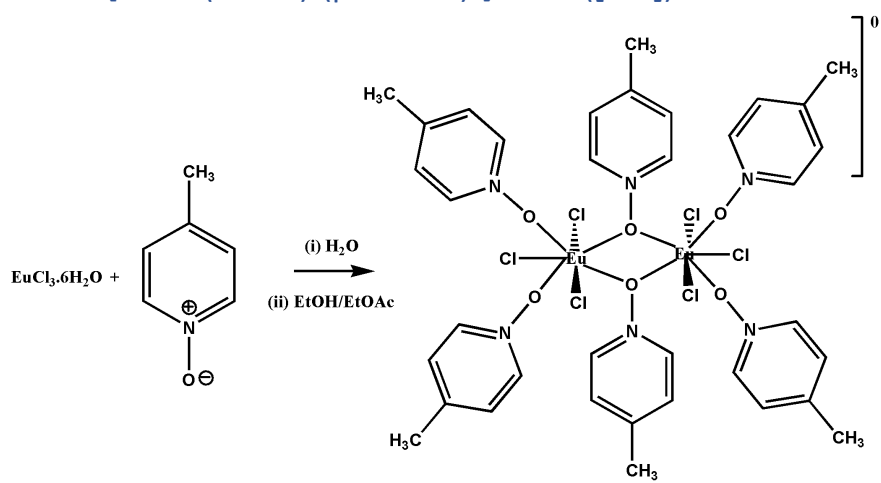

**Scheme S1.** Preparation of the molecular complex **[Eu$_2$]**

To a solution of 0.35 g (3.2 mmol) of 4-picoline N-Oxide in 12 mL of $H_2O$, 1.07 mmol of $EuCl_3·6H_2O$ was added as a solid. The reaction mixture was stirred for 10 min and water was evaporated from the reaction mixture under reduced pressure. The solids were re-dissolved in 15 mL of hot EtOH, filtered, and the filtrate was cooled to RT. About 15-20 mL of EtOAc was carefully added to the filtrate until the formation of slight turbidity/precipitate, and the mixture was filtered to obtain a clear solution. The clear solution was left undisturbed for a few days in a closed vial, yielding X-ray quality crystals of $[Eu_2Cl_6(picNO)_4(\mu_2\text{-}picNO)_2]·2H_2O$ in about 60% yield. The crystalline complex is stable and can be stored and handled at ambient conditions.

**Elemental analysis: $[Eu_2Cl_6(picNO)_4(\mu_2\text{-}picNO)_2]·2H_2O$:** Calculated for $C_{36}H_{42}Cl_6Eu_2N_6O_6·2H_2O$, C 35.8, H 3.84, N 6.96; Found C 35.74, H 3.84, N 6.9.

**S1.3. X-ray crystallography:** X-ray diffraction data of $[Eu_2Cl_6(PicNO)_4(\mu_2\text{-}PicNO)_2]·2H_2O$ was collected on a Bruker APEX II DUO Kappa-CCD diffractometer equipped with an Oxford Cryosystem liquid $N_2$ device, using Mo-K$\alpha$ radiation ($\lambda$ = 0.71073 Å). The crystal-detector distance was 38 mm. The cell parameters were determined (APEX3 software; M86-EXX229V1 APEX3 User Manual", Bruker AXS Inc., Madison, USA, 2016) from reflections taken from three sets of 12 frames, each at 10 s exposure. The structure was solved using the program SHELXT-2014.[1] The refinement and all further calculations were carried out using SHELXL-2014.[2] Hydrogen atoms were included in calculated positions and treated as riding atoms using SHELXL default parameters. The non-hydrogen atoms were refined anisotropically, using weighted full-matrix least-squares on $F^2$.

## S2. Discussion
### S2.1. Packing of [Eu$_2$] in crystalline state

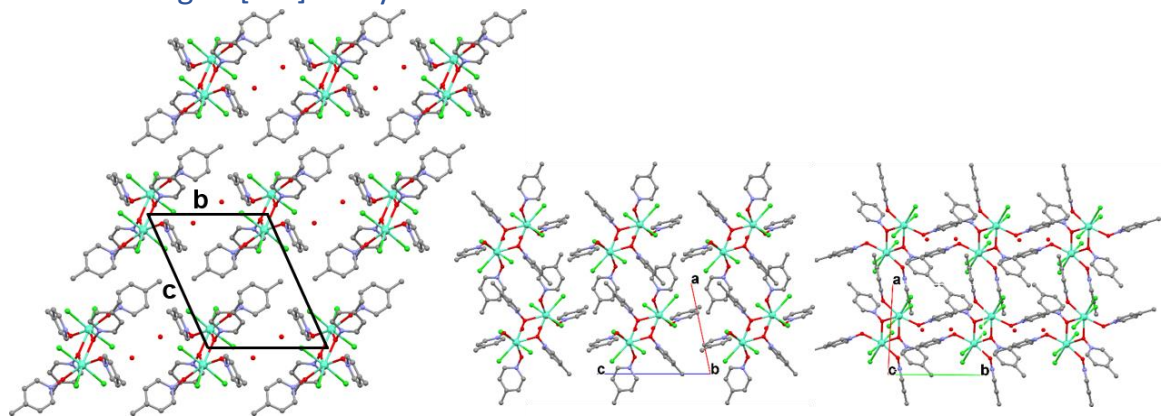

**Fig. S4 |** Arrangement of **[Eu₂]** in crystalline state at 173 K, as obtained from single crystal XRD (hydrogen atoms not displayed). Left: view of the b×c plane, in which molecules arrange into layers with two co-crystallized water molecules, according to oblique *p*1 lattice (1 molecule per lattice). Successive layers superpose with constant lateral shift leading to single-layer periodicity and triclinic structure with 1 molecule per cell.

## S2.2. Comparison between the calculated single crystal and powder X-ray diffraction patterns of [Eu₂]

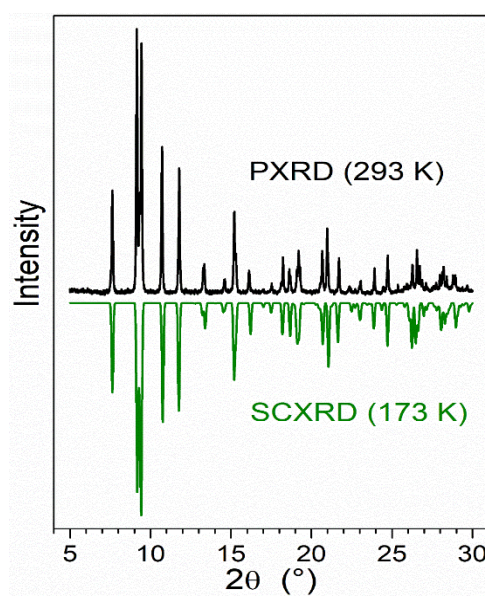

**Fig. S5 |** Comparison between the powder X-ray diffraction (PXRD) and the calculated single crystal X-ray diffraction (SCXRD) pattern (inverted trace at bottom) of **[Eu₂]**. All patterns are composed of same reflections with only tiny changes in peak position and intensity, in relation with tiny variations of lattice parameters and electron densities.

**Table S1.** Structural parameters in single-crystal phase of the complex series at 173 K and bulk crystalline complex at 293 K. Parameters obtained from SCXRD and PXRD data are nearly identical at both temperatures with in a volume expansion at the limit of significance (~0.1%) and an unchanged lateral shift between successive layers (~3.19 Å); the tiny Ö$A/d$ ratio increase (~0.2%) reveals a very slight lateral expansion of layers.

| M | T (K) | Lattice parameters (Å, °), Cell Volume $V$ (Å³) [ $Z$ = 1 molecule par lattice] b×c sublattice area $A$ (Å²), layer spacing $d$ = $d_{100}$ (Å) |
|---|---|---|
| | | |
| SC-XRD | 173 | *a* = 9.7938, *b* = 10.5255, *c* = 12.9179, α = 66.258, β = 75.915, γ = 82.000<br><br>V = 1181.02, A = 124.46, d = 9.4891, ÖA/d = 1.1757 |
| PXRD | 293 | *a* = 9.79(32), *b* = 10.52(56), *c* = 12.90(85), α = 66.57, β = 75.65, γ = 82.29<br><br>V = 1181.(84), A = 124.6(7), d = 9.48, ÖA/d = 1.178 |

## S2.3. Photophysical studies

It is well documented that the $^5D_0 \rightarrow {}^7F_{2,4,6}$ transitions are allowed electric-dipole transitions, which are strongly influenced by the crystal field. Usually, the emission spectra of Eu(III) complexes or more generally Eu(III)-doped phosphors is dominated by the $^5D_0 \rightarrow {}^7F_2$ transition. Examples with intense $^5D_0 \rightarrow {}^7F_4$ are relatively rare and often related to low site-symmetry combined with a highly polarizable chemical environment as observed for 4-picNO-based complex—**[Eu(4-picNO)₈](ClO₄)₃]**.[3] The degree of distortion around Eu(III) is evaluated by measuring the intensity ratio—(I($^5D_0 \rightarrow {}^7F_2$)/I($^5D_0 \rightarrow {}^7F_1$))—called as asymmetry ratio ($R_{21}$). The rationale behind this approach is that the magnetic dipolar $^5D_0 \rightarrow {}^7F_1$ transition is not influenced by the crystal field, whereas the electric dipolar $^5D_0 \rightarrow {}^7F_2$ transition is strongly influenced (hypersensitive) by the local symmetry of the electric/crystal field around a Eu(III) center. The $R_{21}$ = 7.6 obtained for **[Eu₂]** reveals a distorted coordination environment or low-site symmetry around the Eu(III) centers in the complex.

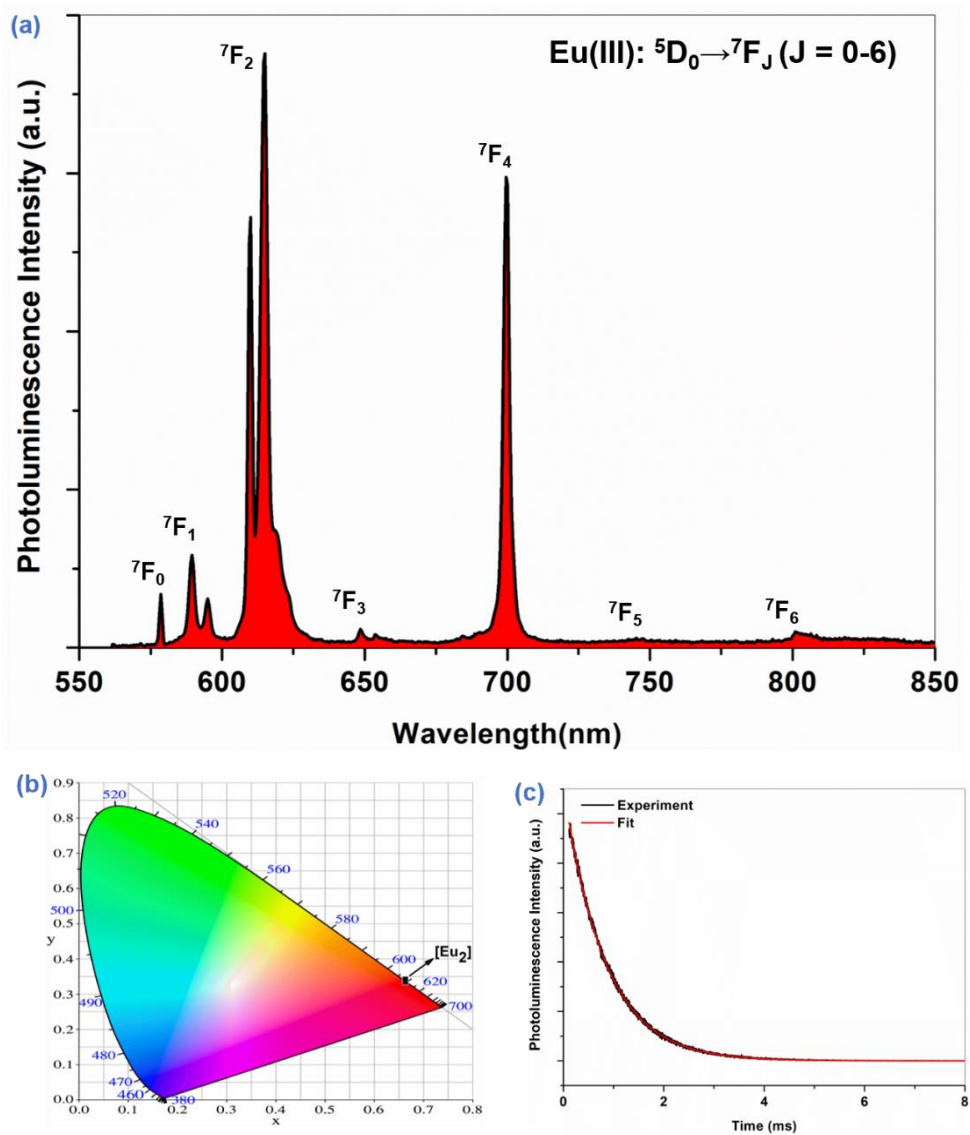

**Fig. S6. (a)** Photoluminescence spectrum showing the $^5D_0 \rightarrow {^7F_J}$ ($J$ = 0-6) transitions of **[Eu₂]** in the visible and near-IR range, **(b)** The CIE chromaticity diagram of **[Eu₂]** in the solid state: x = 0.6615, y = 0.3382, and **(c)** Luminescence decay of the $^5D_0$ excited state of the Eu(III) complex measured at RT, the fluorescence decay is well fitted by a single exponential function, suggesting a single emitting Eu(III) centre.